\documentclass[pdflatex,sn-mathphys-num]{sn-jnl}

\usepackage{graphicx}%
\usepackage{multirow}%
\usepackage{amsmath,amssymb,amsfonts}%
\usepackage{amsthm}%
\usepackage{mathrsfs}%
\usepackage[title]{appendix}%
\usepackage{xcolor}%
\usepackage{textcomp}%
\usepackage{manyfoot}%
\usepackage{float}
\usepackage{soul}
\usepackage{xcolor}
\sethlcolor{yellow}%
\usepackage{listings}%
\usepackage{graphicx}
\usepackage{dcolumn}
\usepackage{bm}
\usepackage{multirow}
\usepackage{amsfonts}
\usepackage{amsthm}
\usepackage{mathrsfs}
\usepackage[title]{appendix}
\usepackage{xcolor}
\usepackage{comment}
\usepackage{hyperref}
\usepackage{textcomp}
\usepackage{braket}
\usepackage{booktabs}
\usepackage{algorithmicx}

\usepackage{algpseudocode}

\makeatletter
\newcounter{algorithm}
\renewcommand{\thealgorithm}{\arabic{algorithm}}
\newenvironment{algorithm}[1][]{%
  \par\medskip\refstepcounter{algorithm}%
  \noindent\begin{minipage}{\linewidth}%
  \hrule height 0.8pt \vskip 2pt
  \renewcommand{\caption}[1]{\textbf{Algorithm \thealgorithm:} ##1\par\vskip 2pt\hrule height 0.4pt\vskip 4pt}%
}{%
  \vskip 2pt\hrule height 0.8pt
  \end{minipage}\par\medskip
}
\makeatother

\theoremstyle{thmstyleone}%
\theoremstyle{thmstyletwo}%

\theoremstyle{thmstylethree}%

\begin{document}

\title[Article Title]{Game, Set, Quantum: Parameterized Quantum Circuit for Correlated  Equilibrium in Bayesian Games}







\author{
Param Pathak$^{1,}$\footnote[1]{These authors contributed equally to this work.},
Vidhi Oad$^{2,*}$,
Nouhaila Innan$^{3,4,}$,
Adarsh Ganesan$^{5,6}$,
Muhammad Shafique$^{3,4}$
}

\affil{\scriptsize{
$^{1}$ QuantumAI Lab, Fractal Analytics, Mumbai, Maharashtra, 40063, India.\\
$^{2}$ Department of Banking, Insurance, and Financial Services, Goa Institute of Management, Sanquelim, Goa, 403505, India.\\
$^{3}$ eBRAIN Lab, Division of Engineering, New York University Abu Dhabi (NYUAD), Abu Dhabi, UAE.\\
$^{4}$ Center for Quantum and Topological Systems (CQTS), NYUAD Research Institute, NYUAD, Abu Dhabi, UAE.\\
$^{5}$ Department of Electrical and Electronics Engineering, Birla Institute of Technology and Science (BITS) Pilani - Dubai Campus, Dubai, 345055, UAE.\\
$^{6}$ Department of Mechanical Engineering, Birla Institute of Technology and Science (BITS) Pilani - Pilani Campus, Rajasthan, 333031, India\\
Emails: param.pathak@fractal.ai, vidhi.ec25@gmail.com, nouhaila.innan@nyu.edu, adarsh@dubai.bits-pilani.ac.in, muhammad.shafique@nyu.edu
}}

\abstract{Strategic decision-making among many agents under incomplete information is central to economics, security, and multi-agent artificial intelligence (AI). Computing equilibria in such settings is challenging because the joint type-action space grows exponentially with the number of players. In binary-type, binary-action Bayesian games with $n$ players, an explicit representation over type-action profiles requires $O(2^{2n})$ entries, making direct linear-programming (LP) formulations increasingly costly as $n$ grows. We propose a hybrid quantum-classical framework for approximating Bayes correlated equilibrium (BCE) using a parameterized quantum circuit (PQC). The PQC represents the conditional distribution over joint actions using $O(nL)$ trainable parameters, where $L$ denotes the circuit depth; for the largest trained setting, $n=8$ and $L=2$, this corresponds to $48$ trainable angles. Each player count is trained independently by maximizing expected social welfare with a penalty on positive aggregated BCE obedience violations. On a strategically coupled Bayesian congestion game with $n=2,4,6,8$ players, feasible PQC solutions attain higher welfare than Monte Carlo counterfactual regret minimization (MCCFR) and discounted counterfactual regret minimization (DCFR) product-strategy baselines while satisfying $\epsilon_{\max}\leq10^{-3}$, where $\epsilon_{\max}$ denotes the maximum positive aggregated BCE obedience violation. Across five independent runs per setting, all runs are feasible for $n=2,4,6$, while four of five are feasible for $n=8$. PQC welfare remains below the exact LP optimum, with the absolute gap increasing with $n$, while classical state-vector simulation prevents PQC training beyond eight players. These results demonstrate the use of a compact PQC parameterization for approximate equilibrium computation and quantify its welfare, feasibility, and classical simulation scaling on the studied benchmark.}




\maketitle

\section{Introduction}\label{sec1}

Many strategic decision-making problems involve agents that act under incomplete information \cite{eisenhardt1992strategic,wang2025research}, where private observations, hidden types, or limited access to other agents' actions shape the resulting equilibria \cite{pappa2015nonlocality,roy2016nonlocal}. In automated markets, security games, distributed optimization, and multi-agent artificial intelligence (AI), each agent observes only part of the state of the interaction and must choose actions while reasoning about hidden information and possible actions of others. Bayesian games provide a standard model for this setting by assigning each player a private type and a utility function that depends on the joint type-action profile \cite{tan1988bayesian,dekel2004learning,zamir2020bayesian}.

\begin{figure}[htbp]
\centering
\includegraphics[width=0.6\linewidth]{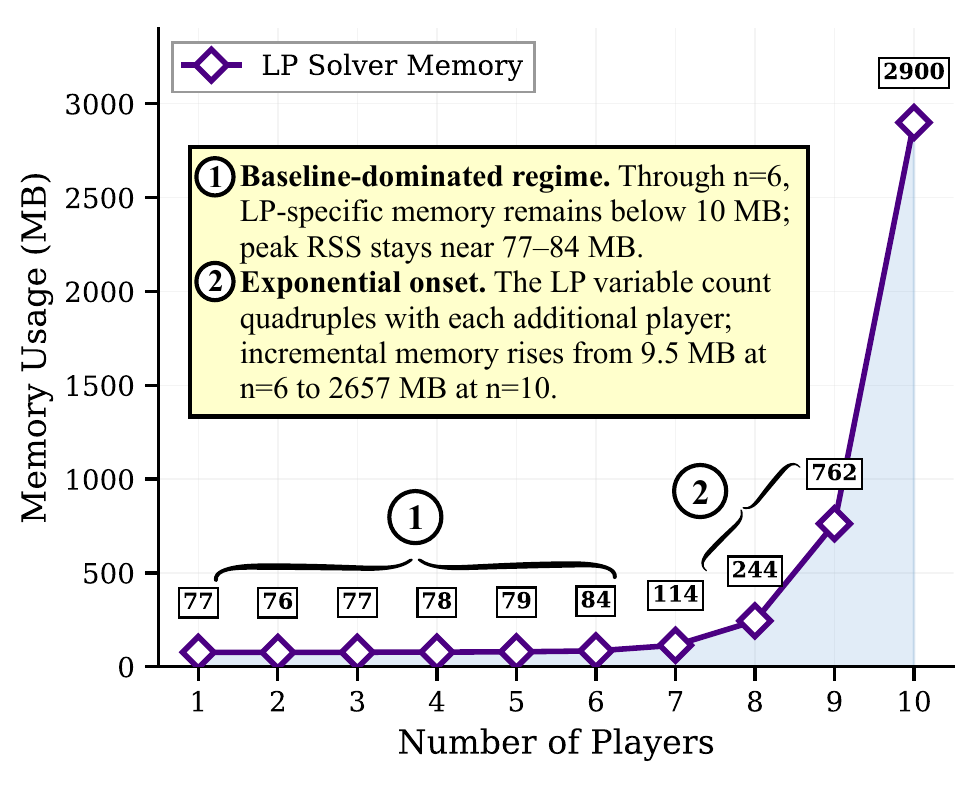}
\caption{Peak memory usage of the direct LP solver across player counts. The flat region below $n=7$ is set by the fixed interpreter and library baseline rather than by the solver, which itself contributes under $10$ MB there. Beyond $n=7$ the LP variable count $4^{n}$ becomes the dominant term, quadrupling with each additional player and reaching $2.9$ GB at $n=10$.}
\label{crash}
\end{figure}

Computing equilibrium strategies in Bayesian games becomes difficult as the number of players grows \cite{van2007monotone}. In the binary-type, binary-action case, the joint type space contains $2^n$ profiles and the joint action space contains $2^n$ profiles, so an explicit distribution over type-action profiles contains $2^{2n}$ entries. Direct linear programming (LP) formulations for Bayes correlated equilibrium computation must operate on this representation \cite{aumann1987correlated,bergemann2016bayes}, leading to rapid growth in memory and runtime. As shown in Fig.~\ref{crash}, the LP reference solver used in this study remains tractable through $n=10$, where the formulation contains $4^{10}=1{,}048{,}576$ type action variables and requires approximately $2.9$ GB of peak memory, illustrating the growing cost of explicit optimization over the distribution.

This bottleneck arises from the explicit representation of the strategy distribution. A scalable solver should describe correlated action recommendations without assigning an independent optimization variable to every type-action probability \cite{bergemann2016information,fujii2025bayes}. Parameterized quantum circuits (PQC) offer one possible compact representation: their measurement probabilities define distributions over bitstrings, while trainable gates control how these distributions depend on encoded inputs \cite{benedetti2019parameterized,du2020expressive}. This structure provides a compact parameterization of conditional strategy distributions in Bayesian games, provided that the induced distributions can be optimized for social welfare while satisfying the obedience conditions required for Bayes correlated equilibrium (BCE).

Based on this motivation, we propose a hybrid quantum-classical framework for approximating Bayes correlated equilibrium in Bayesian games. Instead of storing the conditional strategy distribution, the method represents it with a PQC. The circuit encodes type profiles into a type register and produces the conditional distribution over joint action recommendations from an action register. Each player count is trained independently by gradient based maximization of expected social welfare with a penalty on positive aggregated BCE obedience violations, and reported solutions are selected from checkpoints satisfying the feasibility condition $\epsilon_{\max}\leq10^{-3}$.

The resulting representation uses $\mathcal{O}(nL)$ trainable parameters, where $n$ is the number of players and $L$ is the circuit depth. For the largest PQC setting trained in this study, $n=8$ and $L=2$, this corresponds to $48$ trainable rotation angles. This parameter count is much smaller than the explicit type-action representation, although the method remains a variational approximation and does not guarantee convergence to the welfare optimal Bayes correlated equilibrium for arbitrary games. Extending the same training procedure to $n=10$ requires classical simulation of a $20$-qubit circuit and becomes computationally prohibitive within the available resource budget, so the PQC experiments are limited to eight players; this simulation bottleneck is quantified in the scaling analysis.

The main contributions are as follows:
\begin{itemize}
    \item We formulate approximate Bayes correlated equilibrium computation as a hybrid quantum-classical welfare optimization problem, where a PQC represents the conditional strategy distribution and positive aggregated BCE obedience violations are incorporated through a penalty term.
    
    \item We design a type-conditioned PQC ansatz with $\mathcal{O}(nL)$ trainable parameters and train each player count independently, retaining reported checkpoints according to the equilibrium-feasibility criterion $\epsilon_{\max}\leq10^{-3}$.
    
    \item We evaluate the method on a strategically coupled Bayesian congestion game and compare it with audited MCCFR and DCFR product strategy baselines and an exact LP reference under the same game, utility, and evaluation protocol.
    
    \item We analyze welfare, equilibrium feasibility, the mean clipped deviation metric, runtime, memory usage, circuit depth, and noisy backend sensitivity, identifying where the compact PQC representation is effective and where ansatz expressivity, optimization quality, and classical simulation cost remain limiting factors.
\end{itemize}

The rest of the paper is organized as follows. Section~\ref{sec:background} reviews the relevant background and related work. Section~\ref{method} presents the proposed hybrid quantum-classical BCE optimization framework. Section~\ref{exp} describes the experimental setup. Section~\ref{results} reports the results, and Section~\ref{sec:discussion} discusses limitations and future directions.

\section{Background and related work}\label{sec:background}

\subsection{Bayesian games and Bayes correlated equilibrium}

A finite Bayesian game of incomplete information is defined by
\begin{equation}
\mathcal{G}
=
\left(
N,
\{\Theta_i\}_{i\in N},
\{A_i\}_{i\in N},
\pi,
\{u_i\}_{i\in N}
\right),
\end{equation}
where $N=\{1,\ldots,n\}$ is the set of players, $\Theta_i$ is the private type space of player $i$, $A_i$ is the action space of player $i$, $\pi$ is the common prior over joint type profiles, and $u_i:\Theta \times A \to \mathbb{R}$ is the utility function of player $i$. The joint type and action spaces are
\begin{equation}
\Theta=\prod_{i=1}^{n}\Theta_i,
\qquad
A=\prod_{i=1}^{n}A_i .
\end{equation}
Each player observes only its own type and chooses an action while reasoning about the hidden types and possible actions of the other players.

In the binary-type, binary-action setting studied in this work, $\Theta_i=\{0,1\}$ and $A_i=\{0,1\}$ for every player. Hence, $|\Theta|=2^n$, $|A|=2^n$, and an explicit distribution over type-action profiles contains $|\Theta||A|=2^{2n}$ entries. This exponential growth is the source of the rapid increase in representation size and computational cost faced by direct LP based equilibrium solvers.

Bayes correlated equilibrium extends correlated equilibrium to games with incomplete information. A recommendation rule observes the realized type profile and draws a joint action profile according to a conditional distribution $\sigma(a|\theta)$. The distribution is stable if no player can improve its expected utility by deviating from the recommended action, given its private type and recommendation. In our formulation, this condition is enforced through positive aggregated BCE obedience violations, with their maximum denoted by $\epsilon_{\max}$ and used as the equilibrium feasibility criterion in Sec.~\ref{method}. 

\subsection{Classical regret minimization and equilibrium computation}

Classical approaches to equilibrium computation include exact optimization over equilibrium constraints and iterative regret minimization \cite{farina2019efficient,anagnostides2025computational}. Direct LP formulations can compute equilibria in finite games, but they require explicit access to the type-action distribution and become memory intensive as $|\Theta||A|$ grows~\cite{zhang2020sparsified,im2026scalable}. Regret-minimization methods reduce this burden by updating strategies from deviation incentives rather than solving the LP directly.

Counterfactual regret minimization (CFR) and its variants have been widely used for large imperfect-information games~\cite{zinkevich2007regret}. These methods accumulate regret values associated with information sets and construct average strategies over repeated play. Discounted CFR (DCFR) modifies the update by discounting accumulated regrets and assigning greater weight to later iterations~\cite{xu2024dynamic}. In this work,  Monte Carlo CFR (MCCFR) and DCFR serve as the main classical regret-minimization baselines \cite{lanctot2009monte}. Because their standard convergence guarantees do not directly establish BCE convergence for the general-sum Bayesian game considered here, all methods are evaluated using the same welfare, mean clipped deviation metric $R$, and $\epsilon_{\max}$ metrics.

Related work has also studied scalable correlated-equilibrium concepts in incomplete-information games. Koessler \textit{et al.}~\cite{koessler2024correlated} studied nonatomic Bayesian games and characterized Bayes correlated Wardrop equilibria using action-flow distributions, avoiding explicit finite-player joint strategy representations. Song \textit{et al.}~\cite{song2022sample} developed sample-efficient algorithms for learning extensive-form correlated equilibrium from feedback. These works address scaling through classical representations and sampling, while our approach studies a variational quantum representation of the conditional strategy distribution.

\subsection{Quantum games and variational quantum models}

Quantum game theory studies strategic interaction when game states, strategies, or information structures are represented using quantum systems \cite{lee2003efficiency,gutoski2007toward,khan2018quantum,ikeda2022theory,piispanen2025defining}. Early work by Eisert \textit{et al.}~\cite{eisert1999quantum} showed that quantum strategies can change the equilibrium structure of classical games. Later work examined the computational complexity of equilibria in quantum games. For example, Bostanc{\i} and Watrous~\cite{bostanci2022quantum} showed that approximating Nash equilibria in certain quantum games lies in PPAD, connecting quantum equilibrium computation to classical complexity theory.

Other studies examine quantum decision processes and learning inside quantum games. Bang \textit{et al.}~\cite{bang2016quantum} studied a secret-bit guessing game in which quantum phases enlarge the available reasoning process compared with classical probabilistic strategies. Silva \textit{et al.}~\cite{silva2023maximizing} considered gradient-based learning in quantum games and analyzed the effect of circuit noise on local reward maximization. More recently, Aldridge \textit{et al.}~\cite{aldridge2026projected} represented mixed strategies as PQC Born distributions for approximate Nash equilibrium computation in two-player zero-sum games, while Sezer~\cite{sezer2026quantum} introduced a quantum analogue of Bayes correlated equilibrium based on quantum information structures and semidefinite optimization. These works differ from our setting, where a PQC is used as a computational representation for approximating a classical Bayes correlated equilibrium in a finite Bayesian game.

PQCs provide the main computational tool used in our framework. PQCs have been studied as trainable quantum models with data encoding, variational layers, and measurement-based outputs. The parameter-shift rule provides one standard approach for evaluating gradients of parameterized quantum gates~\cite{mitarai2018quantum,schuld2019evaluating}. Our framework builds on this variational paradigm by using a type-conditioned PQC to represent conditional distributions over joint actions.

\section{Methodology}\label{method}
\begin{figure}[htbp]
    \centering
\includegraphics[width=1\linewidth]{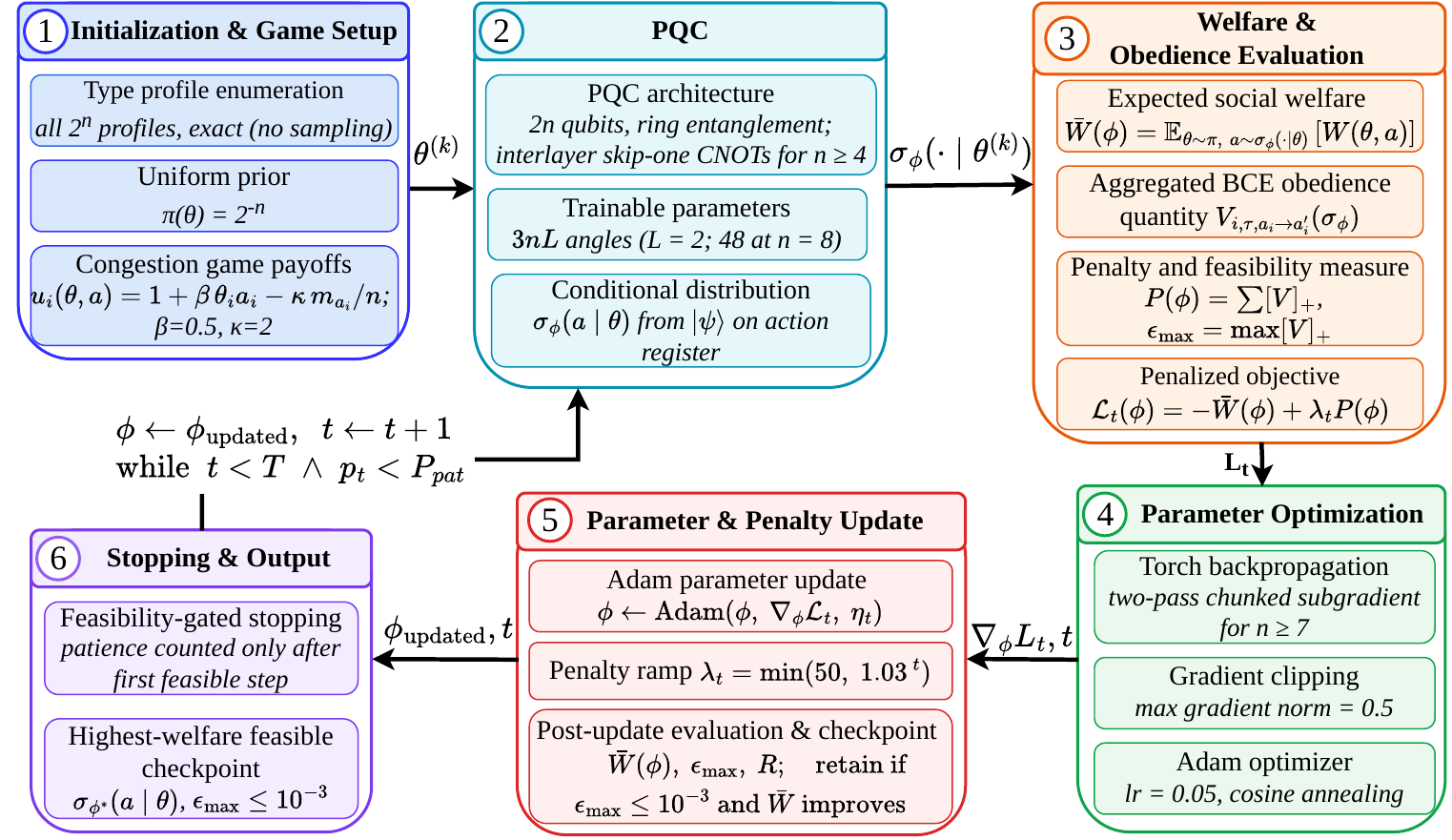}
\caption{Overview of the hybrid quantum-classical workflow for approximating Bayes correlated equilibria, from exact type-profile enumeration and PQC-based strategy generation to welfare and obedience evaluation, parameter optimization, feasibility-gated checkpointing, and final output}.
    \label{QRMA}
\end{figure}

We propose a hybrid quantum-classical method for approximating Bayes correlated equilibrium in Bayesian games. The method represents the conditional action distribution with a PQC, evaluates expected social welfare and positive aggregated BCE obedience violations under the circuit-induced distribution, and updates the circuit parameters to maximize welfare while reducing these violations. As shown in Fig.~\ref{QRMA}, the workflow consists of six stages: type-profile enumeration and game setup; PQC-based generation of the conditional action distribution; welfare and BCE violation evaluation; parameter optimization through automatic differentiation; penalty and checkpoint updates; and feasibility-gated stopping with selection of the highest-welfare checkpoint satisfying $\epsilon_{\max}\leq10^{-3}$.

\subsection{Initialization and type-profile enumeration}
We consider an $n$-player Bayesian game in which each player $i \in \{1,\ldots,n\}$ has a private type $\theta_i \in \Theta_i$ and selects an action $a_i \in A_i$. In this work, we focus on the binary-type, binary-action case, where $\Theta_i=\{0,1\}$ and $A_i=\{0,1\}$ for every player. A joint type profile is denoted by $\theta=(\theta_1,\ldots,\theta_n)$, and a joint action profile is denoted by $a=(a_1,\ldots,a_n)$. The prior over type profiles is uniform and denoted by $\pi(\theta)$, with $\pi(\theta)=2^{-n}$.

Within this framework, we study a Bayesian congestion game. Let $m_x$ denote the number of players that select resource $x$. The payoff to player $i$ is
\begin{equation}
u_i(\theta,a) = 1 + \beta\,\theta_i a_i -\kappa\,\frac{m_{a_i}}{n},
\label{eq:congestion_payoff}
\end{equation}
with $\beta=0.5$ and $\kappa=2$. The second term is a private preference bonus available only to players of type $1$ that select resource $1$, and the third is a congestion cost determined by how many players select the same resource. The payoff of each player therefore depends jointly on its own private type and on the actions of the other players.

Given these payoffs, the target is a conditional strategy distribution $\sigma(a|\theta)$ that approximately satisfies the Bayes correlated equilibrium condition. For every player $i$, type $\theta_i$, recommended action $a_i$, and alternative action $a'_i$, no player should obtain a positive expected gain by deviating unilaterally while the other players follow their recommendations. This condition is expressed as

\begin{equation}
V_{i,\theta_i,a_i,a'_i}(\sigma)
=
\sum_{\theta_{-i},a_{-i}}
\pi(\theta_i,\theta_{-i})
\sigma(a_i,a_{-i}|\theta_i,\theta_{-i})
\left[
u_i(\theta,a'_i,a_{-i})
-
u_i(\theta,a_i,a_{-i})
\right]
\leq 0 ,
\label{eq:bce_condition}
\end{equation}

where $\theta_{-i}$ and $a_{-i}$ denote the types and actions of all players except player $i$. A positive value of $V_{i,\theta_i,a_i,a'_i}$ indicates that the recommendation is not stable for that player, type, and deviation. The left-hand side is weighted by the joint prior and is not divided by $\Pr(\theta_i,a_i)$. Whenever this conditioning event has positive probability, normalization does not alter the sign of the obedience test, while the unnormalized form remains well defined when the event has zero probability.

In our implementation, this condition is evaluated over the type space, and all $2^n$ type profiles are therefore enumerated exactly, with the associated payoff and deviation-gain tables constructed once during initialization rather than resampled during training. No Monte Carlo sampling over type profiles is used at any stage, so the welfare and obedience quantities are evaluated exactly for the current parameters at every iteration. For $n \geq 7$, the gradient is accumulated in chunks across the enumerated profiles in order to bound peak memory; this affects only how the gradient is assembled and leaves the enumeration itself unchanged.

Building on this setup, the circuit is trained independently at each player count $n \in \{2,4,6,8\}$, without curriculum learning or warm-starting across player counts, so the result reported at one player count does not depend on the outcome at any other. Five independent runs with different random seeds are performed at each player count. Circuit parameters are initialized from a zero-mean Gaussian distribution with standard deviation $0.3$, drawn from the seed of the run.

\subsection{Parameterized quantum circuit}

We adopt a hybrid variational quantum-classical framework in which a PQC defines a trainable conditional strategy distribution, while a classical optimizer updates the circuit parameters using gradients of the welfare objective with penalties on positive BCE obedience violations. For each type profile $\theta$, the PQC maps the encoded private information of all players to a probability distribution over joint action profiles. This distribution is then used as the recommendation rule, with its expected social welfare and BCE obedience violations evaluated classically.

\begin{figure}[htbp]
    \centering
    \includegraphics[width=1\linewidth]{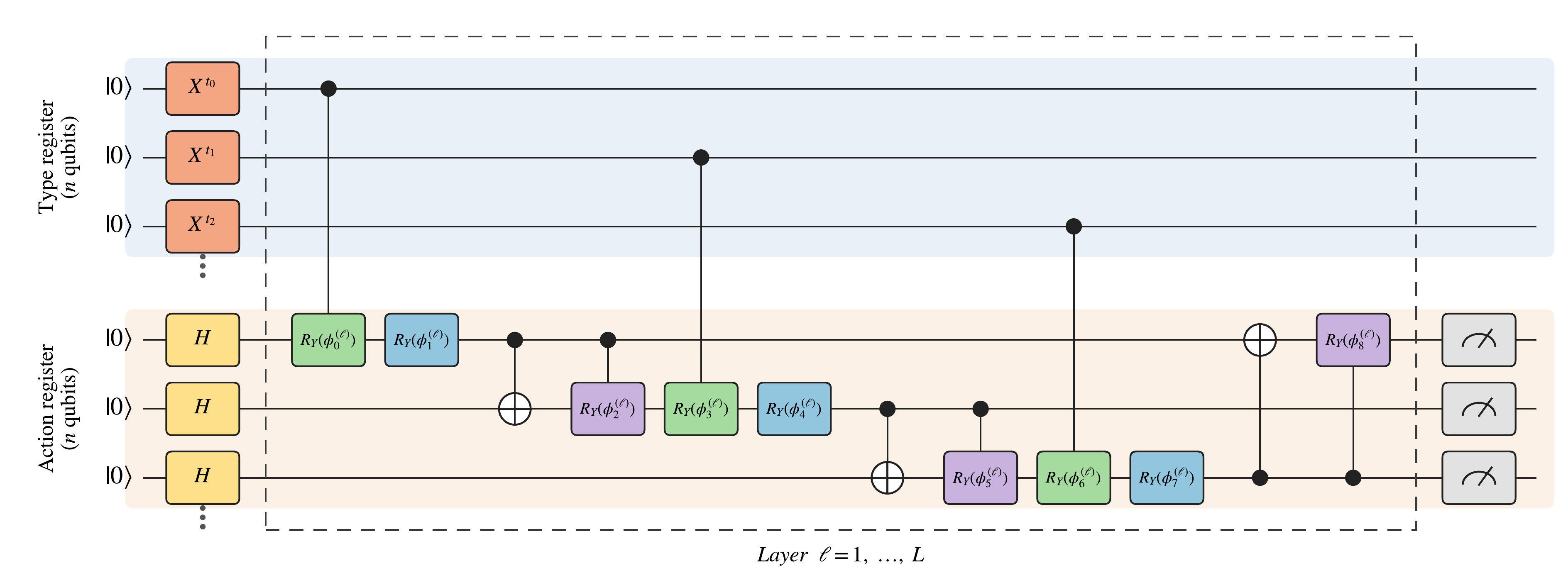}
\caption{Hybrid PQC architecture for approximate Bayes correlated equilibrium computation. The type register encodes private type profiles, while the action register represents the conditional distribution over joint action recommendations. Each layer applies type-conditioned rotations, local action rotations, and ring entangling blocks, yielding $3nL=\mathcal{O}(nL)$ trainable parameters. For $n\geq4$, additional skip-one CNOT connections are applied between consecutive layers; these connections are not active in the $n=3$ schematic shown here.}
    \label{qtmckt}
\end{figure}

As illustrated in Fig.~\ref{qtmckt}, the circuit acts on $2n$ qubits for an $n$-player game. The first $n$ qubits form a type register, and the remaining $n$ qubits form an action register. The type register encodes the current type profile in the computational basis. Specifically, for each player $i$, a Pauli-$X$ gate is applied to the corresponding type qubit when $\theta_i=1$, while the qubit is left in $\ket{0}$ when $\theta_i=0$. These type qubits serve only as control qubits and are not measured. The action register is initialized by Hadamard gates, producing an equal-amplitude superposition over all $2^n$ joint action profiles before the trainable layers are applied.

The ansatz consists of $L$ trainable layers. In each layer $\ell$, and for each player $i$, the circuit applies three parameterized components. First, a type-conditioned controlled rotation $CRY(\phi^{(\ell)}_{3i-2})$ maps information from type qubit $i$ to action qubit $i$, allowing the action distribution to depend on the private type profile. Second, a local trainable rotation $RY(\phi^{(\ell)}_{3i-1})$ is applied to action qubit $i$. Third, neighboring action qubits are coupled through a ring entangling block consisting of a $CNOT$ followed by a controlled rotation $CRY(\phi^{(\ell)}_{3i})$. Between consecutive layers, for $n\geq4$, additional skip-one $CNOT$ gates are applied on the action register from action qubit $i$ to action qubit $(i+2)\bmod n$ for even $i$. These connections increase coupling between non-neighboring action qubits without increasing the number of trainable parameters.

After the final layer, the computational-basis probabilities of the action register are evaluated jointly. During training, the probability vector is obtained directly from the state-vector simulation rather than from sampled action outcomes. These probabilities define the circuit-induced conditional strategy distribution

\begin{equation}
\sigma_\phi(a|\theta)=p_\phi(a|\theta),
\label{eq:circuit_distribution}
\end{equation}

where $\phi$ denotes the set of trainable circuit parameters and $p_\phi(a|\theta)$ is the probability assigned to joint action profile $a$ conditioned on type profile $\theta$.

Each trainable layer contributes three rotation angles per player, giving
\begin{equation}
N_{\mathrm{par}} = 3nL
\label{eq:parameter_count}
\end{equation}
trainable parameters. For the largest PQC configuration trained in this work, $n=8$ and $L=2$, the circuit therefore contains $48$ trainable parameters. This provides a compact variational representation of the conditional strategy distribution compared with a direct tabular representation over all type-action profiles. The compactness, however, comes with an expressivity tradeoff: the circuit does not parameterize the probability simplex, but only the family of distributions reachable by the chosen ansatz.

\subsection{Welfare and obedience evaluation}
\label{sec:objective}

The PQC is trained to identify a high-welfare conditional action distribution while controlling violations of the Bayes correlated equilibrium conditions. For a type profile $\theta$ and joint action profile $a$, let

\begin{equation}
W(\theta,a)
=
\sum_{i=1}^{n} u_i(\theta,a)
\label{eq:social_welfare}
\end{equation}

denote the social welfare, where $u_i(\theta,a)$ is the payoff of player $i$ defined in Eq.~\eqref{eq:congestion_payoff}. The expected welfare induced by the circuit is
\begin{equation}
\overline{W}(\phi)
=
\sum_{\theta}
\pi(\theta)
\sum_{a}
\sigma_{\phi}(a|\theta)
W(\theta,a),
\label{eq:expected_welfare}
\end{equation}
where $\phi$ denotes the trainable circuit parameters, $\pi(\theta)$ is the type-profile prior, and $\sigma_{\phi}(a|\theta)$ is the conditional action distribution generated by the PQC.

Welfare maximization alone does not guarantee equilibrium feasibility. The obedience conditions in Eq.~\eqref{eq:bce_condition} are therefore evaluated when assessing welfare. Since the action space is binary, each recommendation $a_i$ has the unique alternative action $1-a_i$. The penalty is defined as the sum of the positive aggregated obedience violations:

\begin{equation}
P(\phi)
=
\sum_{i=1}^{n}
\sum_{\tau \in \{0,1\}}
\sum_{a_i \in \{0,1\}}
\left[
V_{i,\tau,a_i,1-a_i}
\left(\sigma_{\phi}\right)
\right]_{+},
\label{eq:bce_penalty}
\end{equation}

where $V_{i,\tau,a_i,1-a_i}$ is the joint-prior-weighted BCE obedience quantity defined in Eq.~\eqref{eq:bce_condition}, $\tau$ denotes the private type of player $i$, and $[x]_{+}=\max(0,x)$. Constraints that already satisfy the BCE condition therefore make no contribution to the penalty.

Equilibrium feasibility is assessed directly using the maximum positive aggregated obedience violation

\begin{equation}
\epsilon_{\max}(\phi)
=
\max_{i,\tau,a_i}
\left[
V_{i,\tau,a_i,1-a_i}
\left(\sigma_{\phi}\right)
\right]_{+}.
\label{eq:max_violation}
\end{equation}

A circuit distribution is classified as feasible when

\begin{equation}
\epsilon_{\max}(\phi)\leq 10^{-3}.
\label{eq:feasibility_threshold}
\end{equation}

Since $P(\phi)$ is the sum of all positive aggregated obedience violations and $\epsilon_{\max}(\phi)$ is their maximum,

\begin{equation}
\epsilon_{\max}(\phi)\leq P(\phi).
\label{eq:eps_le_penalty}
\end{equation}

Thus, $P(\phi)\leq10^{-3}$ is sufficient for feasibility under Eq.~\eqref{eq:feasibility_threshold}, although the penalized optimization does not guarantee that $P(\phi)$ reaches this value.

The training objective combines welfare maximization with the obedience penalty. At optimization step $t$, the loss is

\begin{equation}
\mathcal{L}_{t}(\phi)
=
-\overline{W}(\phi)
+
\lambda_t P(\phi),
\label{eq:loss}
\end{equation}

where $\lambda_t$ is the penalty coefficient. Because the loss is minimized, the first term favors higher expected welfare, while the second discourages BCE obedience violations. The penalty coefficient follows

\begin{equation}
\lambda_t
=
\min\left(50,\,1.03^{\,t}\right).
\label{eq:penalty_schedule}
\end{equation}

This schedule begins at $\lambda_0=1$ and progressively increases the weight assigned to BCE feasibility, up to a maximum value of $50$.

In addition to the direct feasibility measure, we retain the mean clipped deviation metric $R$ for evaluation and comparison across the PQC and classical baselines. For a realized type-action profile, define the unilateral deviation gain of player $i$ as

\begin{equation}
g_i(\theta,a)
=
u_i\left(\theta,1-a_i,a_{-i}\right)
-
u_i\left(\theta,a_i,a_{-i}\right).
\label{eq:deviation_gain}
\end{equation}

The mean clipped deviation metric is then

\begin{equation}
R(\phi)
=
\frac{1}{n}
\sum_{i=1}^{n}
\sum_{\theta}
\pi(\theta)
\sum_{a}
\sigma_{\phi}(a|\theta)
\left[g_i(\theta,a)\right]_{+}.
\label{eq:mean_regret}
\end{equation}

$R(\phi)$ is used only as an additional evaluation metric; it is neither minimized during training nor used as the BCE feasibility criterion.

The distinction between $R$ and $\epsilon_{\max}$ is important. The metric $R$ clips realized deviation gains before averaging, whereas the BCE condition first aggregates deviation gains over the relevant type and recommendation events and then tests the resulting obedience quantity. Hence, $R=0$ implies exact BCE obedience, whereas $R>0$ does not by itself imply violation of an aggregated BCE constraint. In the binary-action setting,

\begin{equation}
\epsilon_{\max}(\phi)
\leq
nR(\phi).
\label{eq:regret_epsilon_relation}
\end{equation}

This provides a one-sided relationship between the two metrics, while $\epsilon_{\max}$ remains the quantity used directly to determine equilibrium feasibility.

\begin{algorithm}[h]
\caption{PQC welfare optimization with BCE obedience penalty}
\label{alg:qrm}
\footnotesize
\begin{algorithmic}[1]
\Require Player count $n$, ansatz depth $L$, maximum optimization steps $T$, initial learning rate $\eta_0=0.05$, initial penalty $\lambda_0=1$, penalty growth factor $\gamma=1.03$, maximum penalty $\lambda_{\max}=50$, feasibility tolerance $\epsilon_{\mathrm{tol}}=10^{-3}$, patience limit $P_{\mathrm{pat}}$
\Ensure Selected parameters $\phi^{*}$ and conditional strategy distribution $\sigma_{\phi^{*}}(a|\theta)$

\State Enumerate all $2^n$ type profiles and construct the welfare and deviation gain tables
\State Initialize $\phi$ independently for the current run
\State Initialize Adam with learning rate $\eta_0$ and a cosine learning rate schedule
\State Initialize the best feasible and lowest violation fallback checkpoints as empty, and set the feasible patience counter to zero

\For{$t=0,\ldots,T-1$}
    \State Set $\lambda_t \leftarrow \min(\lambda_{\max},\lambda_0\gamma^t)$
    \State Evaluate $\sigma_{\phi}(a|\theta)$ over all enumerated type profiles
    \State Compute $\overline{W}(\phi)$ and the aggregated BCE obedience violations
    \State Form $\mathcal{L}_t(\phi)=-\overline{W}(\phi)+\lambda_tP(\phi)$

    \If{$n \geq 7$}
        \State Determine the active positive BCE constraints without gradient tracking
        \State Accumulate the corresponding objective subgradient in chunks with the active set fixed
    \Else
        \State Compute $\nabla_{\phi}\mathcal{L}_t$ by automatic differentiation
    \EndIf

    \State Clip the gradient to a maximum global norm of $0.5$
    \State Update $\phi$ using Adam
    \State Update the learning rate using the cosine schedule
    \State Evaluate the updated parameters to obtain $\overline{W}$, $\epsilon_{\max}$, and $R$

    \If{$\epsilon_{\max}\leq\epsilon_{\mathrm{tol}}$}
        \State Record the checkpoint as feasible
        \If{its welfare exceeds the best feasible welfare}
            \State Store the current parameters as the best feasible checkpoint
            \State Reset the feasible patience counter
        \Else
            \State Increment the feasible patience counter
        \EndIf
    \Else
        \State Update the fallback checkpoint if $\epsilon_{\max}$ is the lowest observed
        \If{a feasible checkpoint has already been found}
            \State Increment the feasible patience counter
        \EndIf
    \EndIf

    \If{a feasible checkpoint exists and the patience limit $P_{\mathrm{pat}}$ is reached}
        \State \textbf{break}
    \EndIf
\EndFor

\If{at least one feasible checkpoint was found}
    \State Set $\phi^{*}$ to the highest welfare feasible checkpoint
\Else
    \State Set $\phi^{*}$ to the lowest $\epsilon_{\max}$ checkpoint and mark the run as infeasible
\EndIf

\State \Return $\phi^{*}$ and $\sigma_{\phi^{*}}(a|\theta)=p_{\phi^{*}}(a|\theta)$
\end{algorithmic}
\end{algorithm}
The optimization procedure is summarized in Algorithm~\ref{alg:qrm}. Each run is performed at a fixed player count using the type profiles and payoff tables constructed during initialization. The algorithm optimizes welfare under the obedience penalty and retains the highest-welfare checkpoint satisfying the prescribed BCE tolerance.
\subsection{Parameter optimization}

The circuit parameters are optimized by differentiating the loss in Eq.~\eqref{eq:loss} with respect to $\phi$. Gradients are computed by automatic differentiation through the state-vector simulator. For player counts $n<7$, the loss is evaluated and differentiated in a single backward pass.

For $n\geq7$, evaluating the objective in a single pass requires greater peak memory. We therefore use a two-pass chunked procedure over the enumerated type profiles. In the first pass, the aggregated BCE obedience quantities are evaluated without gradient tracking to identify the positively violated constraints. In the second pass, this active set is held fixed while the objective and its gradient are accumulated across chunks. This procedure changes only the memory organization of the computation and preserves the same objective and corresponding subgradient as the evaluation.

At optimization step $t$, let

\begin{equation}
G_t=\nabla_\phi \mathcal{L}_t(\phi)
\end{equation}

denote the gradient of the training loss with respect to the current circuit parameters. Because the positive-part operator in the obedience penalty introduces non-differentiable points, $G_t$ is interpreted at such points as the subgradient selected by automatic differentiation. For the chunked $n\geq7$ procedure, this subgradient is accumulated with the active set fixed as described above. Before each optimizer update, the gradient is clipped to a maximum global norm of $0.5$:

\begin{equation}
\widetilde{G}_t
=
\begin{cases}
G_t, & \|G_t\|_2 \leq 0.5, \\[4pt]
0.5\,\dfrac{G_t}{\|G_t\|_2}, & \|G_t\|_2 > 0.5.
\end{cases}
\end{equation}

The clipped gradient $\widetilde{G}_t$ is then passed to the Adam optimizer, initialized with learning rate $\eta_0=0.05$. The parameter update is expressed compactly as

\begin{equation}
\phi \leftarrow \operatorname{Adam}\left(\phi,\widetilde{G}_t,\eta_t\right),
\end{equation}

where $\eta_t$ is adjusted using a cosine annealing schedule. Thus, parameter optimization combines automatic differentiation, gradient clipping, Adam updates, and learning-rate annealing, while the BCE penalty coefficient $\lambda_t$ evolves according to Eq.~\eqref{eq:penalty_schedule}.

\subsection{Penalty and checkpoint update}

The BCE penalty coefficient $\lambda_t$ increases according to Eq.~\eqref{eq:penalty_schedule}, progressively placing greater weight on reducing positive obedience violations as training proceeds. The upper bound on $\lambda_t$ prevents the penalty term from growing without limit.

After each parameter update, the circuit is evaluated using the expected welfare $\overline{W}(\phi)$, the maximum BCE violation $\epsilon_{\max}(\phi)$, and the mean clipped deviation metric $R(\phi)$. A checkpoint is classified as feasible when $\epsilon_{\max}(\phi)\leq10^{-3}$. Among all feasible checkpoints encountered during a run, the checkpoint with the highest expected welfare is retained as the best feasible solution.

Each player count is optimized independently, without curriculum learning or parameter transfer across player counts, as described in the initialization procedure.

\subsection{Stopping and output}

Training continues until the maximum number of optimization steps $T$ is reached or the stopping criterion is satisfied. The patience criterion is activated only after the first feasible checkpoint has been identified. Thereafter, training stops when the highest-welfare feasible solution has not improved over the prescribed patience interval. This prevents termination before a distribution satisfying the BCE feasibility criterion has been found.

Among all checkpoints satisfying $\epsilon_{\max}(\phi)\leq10^{-3}$, the one with the highest expected welfare is selected. The final conditional strategy distribution is

\begin{equation}
\sigma_{\phi^{*}}(a|\theta)
=
p_{\phi^{*}}(a|\theta),
\label{eq:final_distribution}
\end{equation}

where $\phi^{*}$ denotes the selected circuit parameters. If at least one feasible checkpoint is found, $\sigma_{\phi^{*}}(a|\theta)$ is reported as the approximate Bayes correlated equilibrium solution.

If no checkpoint satisfies the feasibility threshold within the available optimization steps, the parameters associated with the lowest observed $\epsilon_{\max}$ are retained as a fallback, and the run is marked as infeasible. Equilibrium feasibility is therefore determined directly by $\epsilon_{\max}$ from Eq.~\eqref{eq:max_violation}, while the mean clipped deviation metric $R$ is reported separately as an additional evaluation measure.

\subsection{Design considerations}

The PQC provides a compact variational representation of the conditional strategy distribution, with the parameter count given by Eq.~\eqref{eq:parameter_count}, rather than assigning an independent parameter to every type-action profile. This compactness does not imply that the circuit can represent every possible conditional distribution. The learned strategy is restricted to the family of distributions generated by the chosen circuit architecture and depth.

The scale of this restriction can be quantified directly. A general conditional recommendation rule over $2^n$ type profiles and $2^n$ joint action profiles has $4^n$ entries subject to $2^n$ normalization constraints, leaving $4^n-2^n$ free parameters, whereas the ansatz uses $3nL$ trainable parameters. At $n=8$ and $L=2$, this corresponds to 48 trainable angles compared with $65{,}280$ free parameters for a general conditional recommendation rule. At fixed depth, the trainable parameter count grows linearly with $n$, while the dimension of the unrestricted conditional-distribution space grows exponentially. This comparison shows that the ansatz becomes increasingly restrictive relative to the conditional-distribution space as $n$ grows, but it does not constitute a characterization of the expressive power of the reachable family.

The proposed method should therefore be interpreted as a variational approximation rather than an exact equilibrium solver. Even when a feasible Bayes correlated equilibrium exists, the chosen ansatz may not contain the welfare-optimal equilibrium distribution, and the nonconvex optimization procedure does not guarantee convergence to the best solution representable by the ansatz. These limitations may become more important as the number of players increases while the circuit depth remains fixed.

The focus of this work is the use of a compact quantum-circuit representation for conditional equilibrium distributions and its empirical behavior under common welfare and BCE feasibility criteria. The analysis therefore emphasizes representational compactness and equilibrium quality, while the computational cost of classical simulation is examined separately in the complexity and scaling analysis.

\subsection{Computational complexity and scaling}

The compact parameterization discussed above describes the size of the variational model, but it does not by itself determine the cost of training the circuit on a classical simulator. The PQC uses $2n$ qubits and $3nL$ trainable parameters, so the number of optimized parameters grows linearly with the number of players at fixed circuit depth $L$. In contrast, classical state-vector simulation must explicitly represent the quantum state of all $2n$ qubits.

A state vector for the circuit contains $2^{2n}=4^n$ complex amplitudes. Thus, although the number of trainable parameters is $\mathcal{O}(nL)$, the memory required to represent a single state vector grows exponentially with $n$. In addition, the training objective is evaluated exactly over all $2^n$ type profiles rather than over a sampled subset.

Each circuit evaluation contains $\mathcal{O}(nL)$ parameterized and entangling operations. Combining the state-vector simulation cost with exact enumeration of all type profiles gives an approximate leading-order classical cost for a forward evaluation of the objective of $\mathcal{O}
\left(
nL\,2^{3n}
\right)
=
\mathcal{O}
\left(
nL\,8^n
\right).$

This expression captures the dominant scaling from exact type-profile enumeration and state-vector simulation. Gradient computation through automatic differentiation introduces additional implementation-dependent time and memory overhead, but does not change the exponential dependence on $n$ described above.

For $n=10$ and $L=2$, the circuit contains $20$ qubits and $60$ trainable parameters. Exact evaluation nevertheless requires all $2^{10}=1024$ type profiles, while each state vector contains $2^{20}=1{,}048{,}576$ complex amplitudes. This illustrates the distinction between parameter compactness and classical simulation cost.

For $n\geq7$, the two-pass chunked procedure described above reduces peak memory by processing subsets of the enumerated type profiles separately. Chunking does not change the objective or its corresponding subgradient, and it does not remove the exponential simulation cost.

\section{Experimental setup}\label{exp}

The proposed quantum solver is implemented in \textit{PennyLane}~\cite{bergholm2018pennylane}, with \textit{PyTorch} used for automatic differentiation. The parameterized circuit is constructed as a QNode that maps a type-profile bitstring $\theta \in \{0,1\}^n$ and a parameter vector $\phi \in \mathbb{R}^{3nL}$ to the joint measurement probabilities of the $n$ action qubits. During training, the circuit is simulated using \texttt{default.qubit} with the \textit{PyTorch} interface and backpropagation. For evaluation without gradient tracking, \texttt{lightning.qubit} is used when available.

\subsection{Training protocol and hyperparameters}

The main PQC experiments are conducted independently for player counts $n \in \{2,4,6,8\}$ using circuit depth $L=2$. Five independent runs with seeds $0$ through $4$ are performed at each player count. Each run is initialized separately and trained for at most $T=150$ optimization steps for $n=2$ and $T=300$ steps for $n=4,6,8$, using Adam with initial learning rate $\eta_0=0.05$ and cosine annealing. Gradients are clipped to a maximum global norm of $0.5$. The penalty coefficient is initialized at $\lambda_0=1$, increased by a factor of $1.03$ at each optimization step, and capped at $50$, as defined in Eq.~\eqref{eq:penalty_schedule}. The feasibility tolerance is fixed at $\epsilon_{\max}\leq10^{-3}$, and the patience limit is set to $120$ steps after the first feasible checkpoint is identified.

All $2^n$ type profiles are evaluated exactly during training. For $n\geq7$, the two-pass chunked procedure described in the Methodology is used to reduce peak memory while preserving the same objective and corresponding subgradient. Final parameters are selected from the feasible checkpoints according to the highest expected welfare. If no feasible checkpoint is found, the run is marked as infeasible and the checkpoint with the lowest observed $\epsilon_{\max}$ is retained only for diagnostic reporting.

A depth ablation is additionally performed at $n=4$ using $L=3$ under the same game and evaluation criteria to examine whether increasing circuit depth improves the learned distribution.

\subsection{Computational environment}

All PQC experiments are performed using classical quantum-circuit simulation. Training experiments are conducted for player counts $n\in\{2,4,6,8\}$, while the exact LP, MCCFR, and DCFR baselines are additionally evaluated at $n=10$. Local experiments were conducted on a MacBook Pro equipped with an Apple M4 processor and 16~GB of unified memory. Larger optimization and profiling runs were executed in higher-memory cloud-based CPU environments.

The $n=10$ PQC configuration is evaluated separately as a resource probe rather than through training, because classical simulation of the corresponding $20$-qubit circuit becomes computationally prohibitive within the available resource budget.

\subsection{Benchmark Bayesian game}\label{PF}

All quantum and classical methods are evaluated on the same Bayesian congestion game defined in Eq.~\eqref{eq:congestion_payoff}. Each player has two private types, $\Theta_i=\{0,1\}$, and two available actions, $A_i=\{0,1\}$, interpreted as two competing resources. Player types are drawn independently from the uniform prior, so that $\pi(\theta)=2^{-n}$ for every joint type profile.

For an action profile $a$, let $m_x$ denote the number of players selecting resource $x\in\{0,1\}$. The payoff to player $i$ is
\begin{equation}
u_i(\theta,a)
=
1+\beta\,\theta_i a_i
-
\kappa\,\frac{m_{a_i}}{n},
\label{eq:benchmark_payoff}
\end{equation}

with $\beta=0.5$ and $\kappa=2$. The private bonus rewards a type-$1$ player for selecting resource $1$, while the congestion term reduces payoff as more players select the same resource. The same prior, payoff function, and parameter values are used for the PQC, exact LP, MCCFR, and DCFR experiments, ensuring that all methods are evaluated under identical game conditions.

\subsection{Classical baselines}

To evaluate the PQC against established iterative equilibrium methods on the same Bayesian congestion game, we use implementations of MCCFR and DCFR. Both methods operate over behavioral strategies conditioned on each player's own type and therefore induce product behavioral strategy distributions rather than general correlated recommendation distributions. Their final strategies are evaluated using the expected welfare $\overline{W}$, mean clipped deviation metric $R$, and maximum BCE obedience violation $\epsilon_{\max}$ used for the PQC.

\textbf{MCCFR}~\cite{lanctot2009monte} uses external sampling. At each iteration, a type profile and the actions of the other players are sampled from the current behavioral strategy, while all available actions of the updating player are evaluated. Regret increments are accumulated for each player and private-type information set, and the reported strategy is obtained from the corresponding average behavioral strategy. Five independent runs are performed at each player count using the iteration budgets shown in Table~\ref{tab:mccfr-hyperparams}.

\begin{table}[htbp]
\centering
\caption{Iteration budgets used for the MCCFR baseline.}
\label{tab:mccfr-hyperparams}
\begin{tabular}{cc}
\hline
\textbf{Players ($n$)} & \textbf{Iterations} \\
\hline
2  & 10\,000  \\
4  & 20\,000  \\
6  & 40\,000  \\
8  & 80\,000  \\
10 & 150\,000 \\
\hline
\end{tabular}
\end{table}

\textbf{DCFR}~\cite{xu2024dynamic} uses regret matching with discounted cumulative regrets and a weighted average strategy. The discount parameters are fixed across all player counts at
$\alpha_{\mathrm{D}}=1.5$, $\beta_{\mathrm{D}}=0$, and $\gamma_{\mathrm{D}}=2$, where $\alpha_{\mathrm{D}}$ and $\beta_{\mathrm{D}}$ control the discounting of positive and nonpositive cumulative regrets, respectively, and $\gamma_{\mathrm{D}}$ determines the weighting of later strategies in the average. The subscript $\mathrm{D}$ distinguishes these quantities from the game parameter $\beta$ introduced in Eq.~\eqref{eq:congestion_payoff}. Five independent runs are performed using the iteration budgets listed in Table~\ref{tab:dcfr-hyperparams}.

\begin{table}[htbp]
\centering
\caption{Hyperparameters and iteration budgets used for the DCFR baseline.}
\label{tab:dcfr-hyperparams}
\begin{tabular}{ccccc}
\hline
\textbf{Players ($n$)}
& $\boldsymbol{\alpha_{\mathrm{D}}}$
& $\boldsymbol{\beta_{\mathrm{D}}}$
& $\boldsymbol{\gamma_{\mathrm{D}}}$
& \textbf{Iterations} \\
\hline
2  & 1.50 & 0.00 & 2.00 & 8\,000   \\
4  & 1.50 & 0.00 & 2.00 & 15\,000  \\
6  & 1.50 & 0.00 & 2.00 & 30\,000  \\
8  & 1.50 & 0.00 & 2.00 & 60\,000  \\
10 & 1.50 & 0.00 & 2.00 & 120\,000 \\
\hline
\end{tabular}
\end{table}

For both baselines, the average product behavioral strategy obtained after the prescribed iteration budget is evaluated under the exact uniform type prior. Expected welfare, $R$, and $\epsilon_{\max}$ are computed using the common game parameters and metric definitions, ensuring a consistent evaluation protocol across all methods.

\section{Results}\label{results}

\subsection{Evaluation metrics}

The PQC and classical baselines are evaluated using a common set of metrics for solution quality, equilibrium feasibility, and computational cost. For all methods, welfare and obedience quantities are computed on the same Bayesian congestion game under the same uniform prior over type profiles.

\subsubsection{Expected social welfare.}

Solution quality is measured by the expected social welfare $\overline{W}$ defined in Eq.~\eqref{eq:expected_welfare}. Higher values indicate greater total expected utility across players. Welfare is interpreted in relation to BCE feasibility rather than independently of the obedience constraints.

\subsubsection{BCE feasibility.}

Equilibrium feasibility is measured by the maximum positive aggregated obedience violation $\epsilon_{\max}$ defined in Eq.~\eqref{eq:max_violation}. A solution is classified as feasible when $\epsilon_{\max}\leq10^{-3}$, as specified in Eq.~\eqref{eq:feasibility_threshold}. For experiments with multiple random seeds, we also report the number of runs satisfying this criterion.

\subsubsection{Mean clipped deviation metric.}

The mean clipped deviation metric $R$ defined in Eq.~\eqref{eq:mean_regret} is retained as an additional measure of unilateral deviation incentives. Lower values indicate smaller positive realized deviation gains on average. The metric is not used as the BCE feasibility criterion, since $R>0$ does not necessarily imply violation of an aggregated BCE obedience constraint.

\subsubsection{Runtime.}

Runtime is measured as wall-clock time for a solver run at a fixed player count. For the PQC, this includes circuit evaluation, automatic differentiation, parameter optimization, and checkpoint evaluation. Runtime is reported separately for each player count because each PQC configuration is trained independently.

\subsubsection{Memory usage.}

Memory usage is measured using peak resident set size (RSS), defined as the maximum process memory observed during a solver run. For the PQC, this includes state-vector simulation and automatic-differentiation memory; for MCCFR and DCFR, it includes the behavioral-strategy and regret-accumulation structures used during optimization. Parameter count is reported separately to characterize the size of the learned representation, while peak RSS captures the memory required during execution.

\subsection{Mean clipped deviation metric across player counts}

We first assess unilateral deviation incentives using the mean clipped deviation metric $R$ defined in Eq.~\eqref{eq:mean_regret}. The metric is evaluated over the joint type space for each tested player count using the same benchmark utility and payoff construction for all methods. Lower values indicate smaller positive deviation gains on average.

\begin{table}[htbp]
\centering
\caption{Mean clipped deviation metric across player counts on the matched utility benchmark. Lower values indicate smaller positive deviation gains on average.}
\label{tab:regret_comparison}
\begin{tabular}{cccc}
\hline\hline
\textbf{Players ($n$)} & \textbf{Quantum (ours)} & \textbf{MCCFR} & \textbf{DCFR} \\
\hline
2  & \textbf{0.000} & 0.003          & 0.375 \\
4  & \textbf{0.000} & 0.156          & 0.250 \\
6  & \textbf{0.045} & 0.195          & 0.214 \\
8  & \textbf{0.049} & 0.167          & 0.183 \\
10 & --             & \textbf{0.136} & 0.158 \\
\hline\hline
\end{tabular}
\end{table}

As shown in Table~\ref{tab:regret_comparison}, the PQC achieves the lowest mean clipped deviation metric at all evaluated PQC player counts, with values remaining below $0.05$ through $n=8$. The PQC values are aggregated over feasible independent runs: all five runs at $n=2$, $4$, and $6$, and the four feasible runs at $n=8$. MCCFR increases from $0.003$ at $n=2$ to $0.195$ at $n=6$ before decreasing to $0.136$ at $n=10$, whereas DCFR decreases from $0.375$ at $n=2$ to $0.158$ at $n=10$. At the matched player counts $n=2$, $4$, $6$, and $8$, both classical baselines remain above the PQC. At $n=10$, the PQC configuration was evaluated only through resource profiling, so a final mean clipped deviation metric is not reported.

\begin{figure}[htbp]
    \centering
    \includegraphics[width=0.6\linewidth]{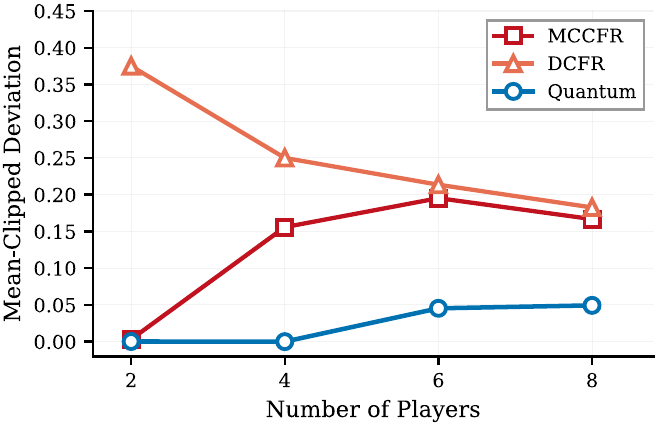}
\caption{Mean clipped deviation metric for the PQC, MCCFR, and DCFR on the matched utility benchmark. The PQC remains below both classical baselines at all evaluated PQC player counts.}
    \label{fig:comparison_all3}
\end{figure}

The same pattern is illustrated in Fig.~\ref{fig:comparison_all3}, which shows the variation in $R$ across methods and player counts. Although the separation between methods varies with $n$, the PQC consistently exhibits the lowest $R$ across the evaluated PQC player counts. These results should be considered in relation to the BCE feasibility analysis, since $R$ is an evaluation metric rather than the direct feasibility criterion.

\subsection{Welfare performance across player counts}

The mean clipped deviation metric measures unilateral deviation incentives but does not directly quantify the welfare generated by the resulting strategy. We therefore evaluate expected social welfare on the same benchmark using the same type distribution and payoff structure.

\begin{figure}[htbp]
\centering
\includegraphics[width=0.6\linewidth]{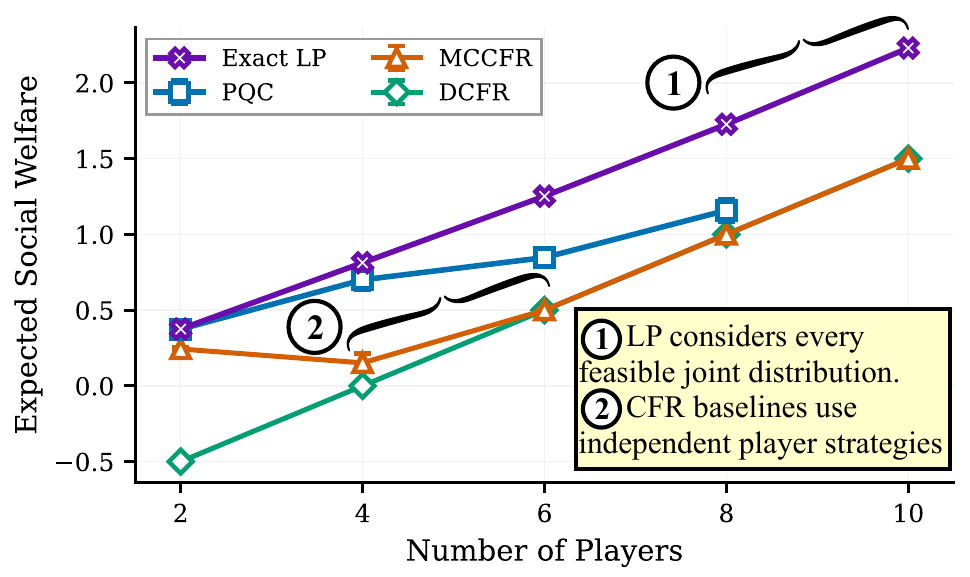}
\caption{Expected social welfare of the exact LP solution, PQC, MCCFR, and DCFR on the matched utility benchmark. The exact LP achieves the highest welfare across all tested player counts, while the PQC remains above both classical baselines at all evaluated PQC player counts.}
\label{fig:welper}
\end{figure}

As shown in Fig.~\ref{fig:welper}, the exact LP achieves the highest expected social welfare across all tested player counts, increasing from approximately $0.38$ at $n=2$ to $2.23$ at $n=10$. At the evaluated PQC player counts, the PQC remains below the LP optimum while exceeding MCCFR and DCFR at $n=2$, $4$, $6$, and $8$. The gap between the PQC and LP increases with $n$, although the experiments do not isolate whether this behavior arises from ansatz expressivity, nonconvex optimization, or both. MCCFR and DCFR optimize regret-based objectives rather than welfare directly, so lower deviation incentives do not necessarily correspond to higher social welfare.

\begin{table}[htbp]
\centering
\caption{Expected social welfare of the PQC across independent seeds, with the exact LP optimum for reference. Values are reported as mean $\pm$ standard deviation over seeds satisfying $\epsilon_{\max}\leq10^{-3}$. All five seeds are feasible at $n=2$, $4$, and $6$, while four of five are feasible at $n=8$.}
\label{tab:welfare_seeds}
\begin{tabular}{cccc}
\toprule
Players ($n$) & Feasible seeds & PQC welfare (mean $\pm$ std) & Exact LP \\
\midrule
2 & 5/5 & $0.3743 \pm 0.0005$ & 0.3750 \\
4 & 5/5 & $0.7000 \pm 0.0658$ & 0.8125 \\
6 & 5/5 & $0.8468 \pm 0.0426$ & 1.2521 \\
8 & 4/5 & $1.1587 \pm 0.0742$ & 1.7266 \\
\bottomrule
\end{tabular}
\end{table}

Table~\ref{tab:welfare_seeds} summarizes the PQC welfare across independent seeds. Variability is greater at the larger tested player counts than at $n=2$. At $n=8$, the reported statistics are computed over the four feasible seeds because one run does not satisfy the BCE feasibility tolerance.
\subsection{BCE feasibility across player counts}
The welfare comparison above measures the quality of the resulting joint behavior, but high welfare alone does not establish that the strategy satisfies the BCE obedience constraints. We therefore assess feasibility using the maximum positive aggregated BCE obedience violation $\epsilon_{\max}$, with a run classified as feasible when $\epsilon_{\max}\leq10^{-3}$.

\begin{table}[htbp]
\centering
\caption{BCE feasibility across player counts for the PQC and the two classical product-strategy baselines. A run is classified as feasible when $\epsilon_{\max}\leq10^{-3}$, and mean $\epsilon_{\max}$ is reported over feasible runs only. The PQC was not trained at $n=10$.}
\label{tab:bce_feasibility}
\begin{tabular}{ccccccc}
\hline\hline
& \multicolumn{2}{c}{\textbf{PQC (ours)}}
& \multicolumn{2}{c}{\textbf{MCCFR}}
& \multicolumn{2}{c}{\textbf{DCFR}} \\
\cmidrule(lr){2-3}\cmidrule(lr){4-5}\cmidrule(lr){6-7}
\textbf{Players ($n$)}
& Feasible & Mean $\boldsymbol{\epsilon_{\max}}$
& Feasible & Mean $\boldsymbol{\epsilon_{\max}}$
& Feasible & Mean $\boldsymbol{\epsilon_{\max}}$ \\
\hline
2  & 5/5 & $1.4\times10^{-6}$ & 4/5 & $3.0\times10^{-4}$ & 5/5 & $7.3\times10^{-13}$ \\
4  & 5/5 & $4.3\times10^{-7}$ & 4/5 & $2.0\times10^{-4}$ & 5/5 & $1.6\times10^{-12}$ \\
6  & 5/5 & $5.1\times10^{-6}$ & 4/5 & $7.0\times10^{-5}$ & 5/5 & $3.6\times10^{-13}$ \\
8  & 4/5 & $5.1\times10^{-4}$ & 5/5 & $8.7\times10^{-5}$ & 5/5 & $5.6\times10^{-14}$ \\
10 & --  & --                 & 5/5 & $2.4\times10^{-5}$ & 5/5 & $8.2\times10^{-15}$ \\
\hline\hline
\end{tabular}
\end{table}

As shown in Table~\ref{tab:bce_feasibility}, all five PQC seeds satisfy the BCE tolerance at $n=2$, $4$, and $6$, with mean $\epsilon_{\max}$ on the order of $10^{-6}$ or smaller. At $n=8$, four of five seeds are feasible, with mean $\epsilon_{\max}=5.09\times10^{-4}$, while the remaining seed reaches a lowest observed value of $\epsilon_{\max}=3.02\times10^{-3}$. This indicates less consistent feasibility at the largest trained player count. Accordingly, the welfare and mean clipped deviation metric $R$ at $n=8$ are aggregated over the four feasible seeds only.

The classical baselines exhibit a different pattern under the same criterion. MCCFR misses the tolerance in one seed at $n=2$, $4$, and $6$, while all five seeds are feasible at $n=8$ and $10$ under the tested iteration budgets. DCFR satisfies the tolerance in every seed at every tested player count, with mean $\epsilon_{\max}$ between approximately $10^{-12}$ and $10^{-15}$. These small aggregated obedience violations can coexist with positive values of $R$, reflecting the distinction between aggregated BCE feasibility and realized deviation gains. Feasibility and welfare are therefore reported separately, particularly because MCCFR and DCFR are restricted to product behavioral strategies, whereas the PQC represents correlated conditional distributions.

\subsection{Runtime and memory usage}\label{mryres}

We next analyze computational cost in terms of peak memory usage and wall-clock runtime. The comparison distinguishes the size of the learned representation from the memory required during computation. This distinction is particularly relevant for the PQC, which uses a compact parameterization while requiring substantially greater memory and runtime under classical state-vector simulation.

\begin{figure}[htbp]
\centering
\includegraphics[width=0.6\linewidth]{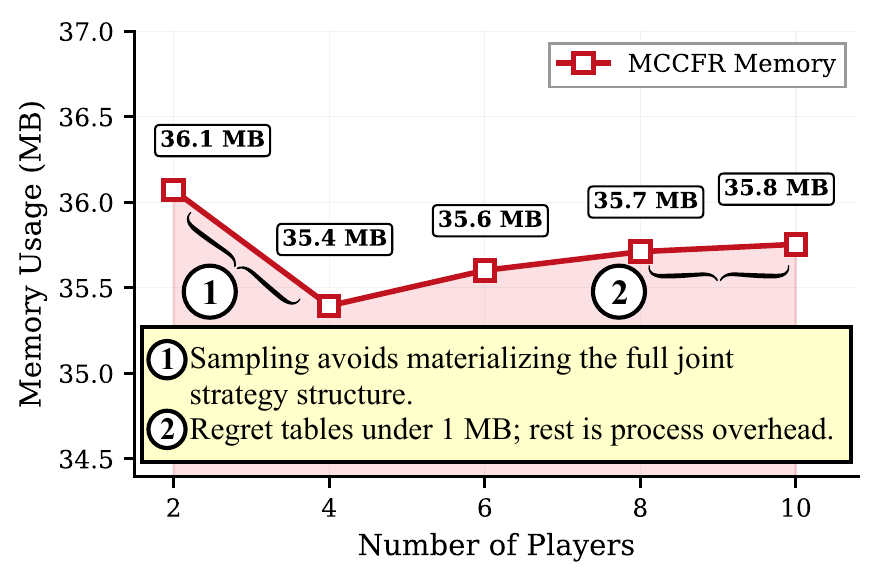}
\caption{Peak memory usage of MCCFR across player counts. MCCFR stores regret and strategy tables per information set, with storage scaling as $O(n|\Theta_i||A_i|)$. At $n=10$, each table contains $40$ action entries, and the solver-specific allocation remains below $1$ MB across the tested player counts.}
\label{fig:mccfr_memory}
\end{figure}

The direct LP reference exhibits a different memory profile. As shown in Fig.~\ref{crash}, peak memory remains approximately $77$--$84$ MB through $n=6$ before increasing to $2.9$ GB at $n=10$. At smaller player counts, the measurements are dominated by a fixed process baseline of approximately $74$ MB, while the LP-specific allocation remains below $10$ MB. As $n$ increases, the explicit type-action variable count scales as $4^n$. The LP contains $4n$ BCE obedience inequalities and $2^n$ conditional normalization equalities; at $n=10$, these correspond to $40$ obedience inequalities and $1024$ normalization equalities, compared with $1{,}048{,}576$ variables. Runtime increases from approximately $0.14$ s at $n=6$ to $666$ s at $n=10$.

MCCFR remains within approximately $35.4$--$36.1$ MB across the tested player counts, as shown in Fig.~\ref{fig:mccfr_memory}. Its regret and strategy tables scale as $O(n|\Theta_i||A_i|)$ rather than with the joint type-action space, amounting to $40$ action entries per table at $n=10$. External sampling avoids explicitly materializing the joint distribution during each iteration. The solver-specific allocation remains below $1$ MB, while runtime increases from approximately $2.0$ s at $n=2$ to $80.7$ s at $n=10$.

\begin{figure}[htbp]
\centering
\includegraphics[width=0.6\linewidth]{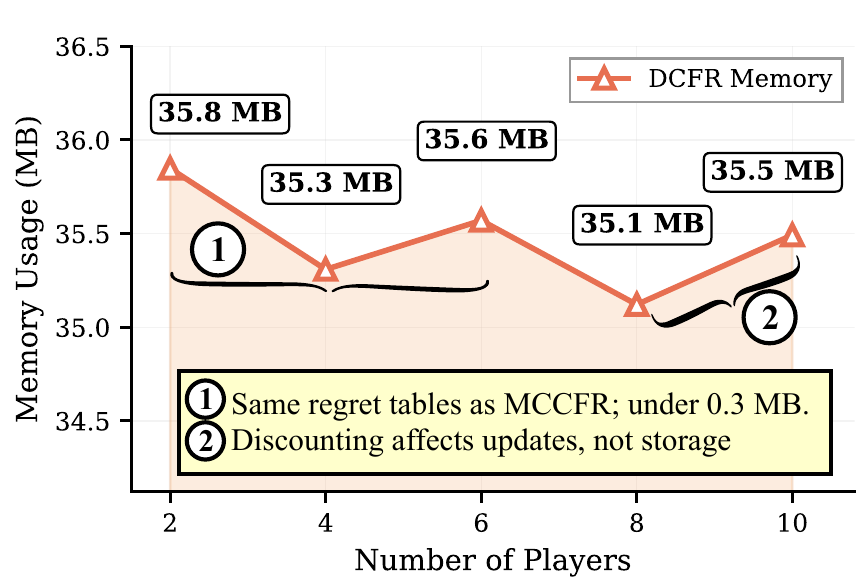}
\caption{Peak memory usage of DCFR across player counts. DCFR uses the same per-information-set regret and strategy-table structure as MCCFR, giving $O(n|\Theta_i||A_i|)$ storage. Its solver-specific allocation remains below approximately $0.3$ MB across the tested player counts.}
\label{fig:dcfr_memory}
\end{figure}

DCFR shows a similar memory profile, remaining within approximately $35.1$--$35.8$ MB across the tested player counts, as shown in Fig.~\ref{fig:dcfr_memory}. Discounting changes how cumulative regrets and average strategies are updated but does not change the size of the stored tables. At $n=10$, the measured MCCFR and DCFR memory footprints differ by approximately $0.26$ MB.

\begin{figure}[htbp]
\centering
\includegraphics[width=0.6\linewidth]{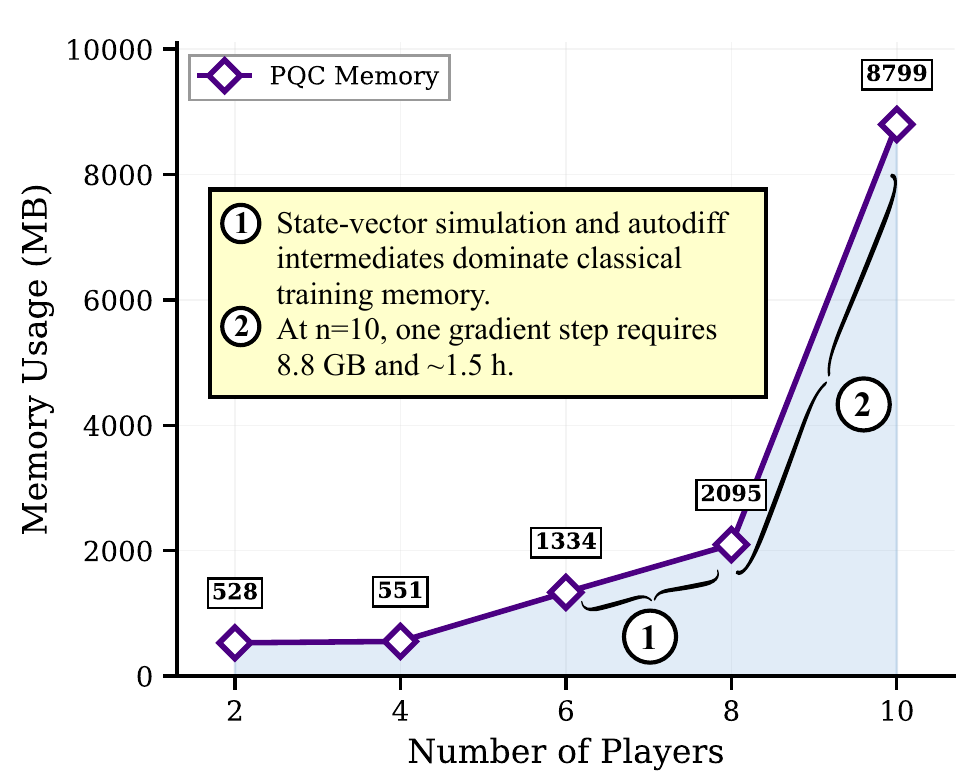}
\caption{Peak memory usage of the classically simulated PQC across player counts. Although the representation contains only $3nL$ trainable parameters, classical state-vector simulation and automatic differentiation produce substantially larger memory requirements. Chunked gradient accumulation reduces peak memory for $n\geq7$, while the $n=10$ resource probe reaches approximately $8.8$ GB.}
\label{fig:pqc_memory}
\end{figure}

The PQC exhibits a different relationship between representation size and execution memory. At $n=8$ and $L=2$, the circuit contains $48$ trainable parameters, compared with $65{,}536$ entries in an explicit type-action representation over $2^{2n}$ profiles. However, this compact parameterization does not translate into low memory usage under classical simulation. As shown in Fig.~\ref{fig:pqc_memory}, peak memory increases from approximately $528$ MB at $n=2$ to $2095$ MB at $n=8$, while the $n=10$ resource probe reaches approximately $8.8$ GB for a single gradient step. The dominant costs arise from state-vector simulation and automatic differentiation; chunked accumulation for $n\geq7$ reduces peak memory but does not eliminate these computational costs. At $n=10$, the PQC resource probe therefore uses more peak memory than the exact LP, approximately $8.8$ GB compared with $2.9$ GB.

Runtime exhibits a similar pattern. At $n=10$, MCCFR and DCFR require approximately $81$ s and $60$ s, respectively, averaged over five runs. PQC training requires approximately $6.3$ hours per run at $n=8$, while a single gradient step in the $n=10$ resource probe requires approximately $1.5$ hours. These measurements reflect the cost of repeated state-vector simulation and gradient evaluation.

These results distinguish the compact size of the PQC parameterization from the substantially larger computational cost of training it through classical state-vector simulation.

\subsection{Ablation: circuit depth}\label{ablation}

The circuit depth $L$ determines the trainable parameter count $3nL$ and is a central architectural hyperparameter. To assess the effect of depth, we compare $L=2$ and $L=3$ at $n=4$ while holding the training protocol fixed: the same $300$-step budget, learning-rate and penalty schedules, $\epsilon_{\max}\leq10^{-3}$ selection criterion, and five random seeds. The player count $n=4$ is used because it allows both circuit depths to be evaluated across all five seeds while exhibiting a non-negligible welfare gap to the exact LP reference.

\begin{table}[htbp]
\centering
\caption{Effect of circuit depth at $n=4$ over five seeds, with the exact LP optimum at $0.8125$. Increasing $L$ from $2$ to $3$ raises the parameter count from $24$ to $36$, while mean welfare remains nearly unchanged.}
\label{tab:depth_ablation}
\begin{tabular}{lcc}
\hline\hline
& $\boldsymbol{L=2}$ & $\boldsymbol{L=3}$ \\
\hline
Trainable parameters $3nL$          & $24$                  & $36$                  \\
Welfare (mean $\pm$ std)            & $0.7000 \pm 0.0658$   & $0.6990 \pm 0.0396$   \\
Gap to LP optimum                   & $0.1125$              & $0.1135$              \\
Feasible seeds                      & $5/5$                 & $5/5$                 \\
Mean $\epsilon_{\max}$              & $4.3\times10^{-7}$    & $1.4\times10^{-9}$    \\
First feasible step (mean; range)   & $24$ ($10$--$42$)     & $18$ ($15$--$21$)     \\
Mean runtime per seed               & $205$ s               & $302$ s               \\
\hline\hline
\end{tabular}
\end{table}

As shown in Table~\ref{tab:depth_ablation}, increasing the depth from $L=2$ to $L=3$ does not improve mean welfare. The corresponding values are $0.7000 \pm 0.0658$ and $0.6990 \pm 0.0396$, while the gaps to the exact LP optimum are $0.1125$ and $0.1135$, respectively. All five seeds satisfy the BCE feasibility tolerance at both depths.

Increasing the circuit depth changes the optimization behavior. At $L=3$, the welfare spread across seeds is reduced and feasibility is reached earlier and more consistently, with the first feasible checkpoint occurring at step $18$ on average compared with step $24$ for $L=2$. These changes do not lead to higher mean welfare.

The additional layer increases the parameter count by $50\%$ and the mean runtime per seed from $205$ s to $302$ s. We therefore retain $L=2$ for the main experiments. Keeping $L$ fixed also preserves the interpretation of the parameter count as $O(nL)$ when comparing scaling across player counts. The residual gap to the exact LP optimum remains after increasing the depth from $L=2$ to $L=3$, indicating that additional circuit depth alone does not account for the observed welfare shortfall at this problem size.

\subsection{Hardware-calibrated noisy backend simulation}\label{hwval}

To assess the sensitivity of the trained circuit to realistic device noise, we evaluate the trained parameters on two IBM Heron calibrated noise-model backends: \texttt{FakeTorino} and \texttt{FakeMarrakesh}~\cite{javadi2024quantum}. These backends emulate device-specific connectivity, gate errors, and noise characteristics, allowing the trained parameters to be evaluated under hardware-calibrated conditions without execution on a physical quantum processor. We consider $n=2$ and $n=4$, for which noisy simulation remains computationally tractable.

For each player count, the circuit is trained on the noiseless simulator and the resulting parameters are then evaluated on the calibrated noise-model backend using $4096$ shots per type profile. For each seed, the same parameter vector is used for the noiseless and noisy evaluations, so the observed differences reflect both calibrated backend noise and finite-shot sampling.

\begin{table}[htbp]
\centering
\caption{Hardware-calibrated noisy-backend simulation on IBM Heron noise models. Welfare and $\epsilon_{\max}$ are reported as means over five trained seeds. Noisy evaluations use $4096$ shots per type profile, and feasibility is assessed using $\epsilon_{\max}\leq10^{-3}$.}
\label{tab:hw_validation}
\renewcommand{\arraystretch}{1.25}
\setlength{\tabcolsep}{7pt}
\begin{tabular}{lcccccc}
\toprule
& & & \multicolumn{2}{c}{\textbf{Welfare}} & \multicolumn{2}{c}{$\boldsymbol{\epsilon_{\max}}$}\\
\cmidrule(lr){4-5}\cmidrule(lr){6-7}
\textbf{Backend} & \textbf{Processor} & $\boldsymbol{n}$ & Noiseless & Noisy & Noiseless & Noisy\\
\midrule
\multirow{2}{*}{\texttt{FakeTorino}}
& \multirow{2}{*}{\textit{Heron r1} (133 qubits)}
& 2 & $0.374$ & $0.287$ & $1.4\times10^{-6}$ & $1.2\times10^{-2}$ \\
& & 4 & $0.700$ & $0.448$ & $4.3\times10^{-7}$ & $3.1\times10^{-2}$\\
\midrule
\multirow{2}{*}{\texttt{FakeMarrakesh}}
& \multirow{2}{*}{\textit{Heron r2} (156 qubits)}
& 2 & $0.374$ & $0.335$ & $1.4\times10^{-6}$ & $5.1\times10^{-3}$ \\
& & 4 & $0.700$ & $0.538$ & $4.3\times10^{-7}$ & $2.3\times10^{-2}$ \\
\bottomrule
\end{tabular}
\end{table}

As shown in Table~\ref{tab:hw_validation}, welfare decreases under noisy evaluation for both backends and both player counts. The absolute reductions are $0.087$ and $0.252$ on \texttt{FakeTorino} at $n=2$ and $n=4$, respectively, and $0.039$ and $0.162$ on \texttt{FakeMarrakesh}. The degradation is larger at $n=4$ for both backends, while \texttt{FakeMarrakesh} exhibits smaller welfare reductions than \texttt{FakeTorino} at both tested player counts.

The effect on BCE feasibility is more pronounced. Noiseless $\epsilon_{\max}$ is at most $1.4\times10^{-6}$, whereas under noisy evaluation it increases to values between $5.1\times10^{-3}$ and $3.1\times10^{-2}$. No seed satisfies the $\epsilon_{\max}\leq10^{-3}$ tolerance on either backend at either player count. Finite-shot sampling also contributes uncertainty: with $4096$ shots, the standard error of an individual estimated action probability is at most approximately $7.8\times10^{-3}$. The observed increase in $\epsilon_{\max}$ therefore reflects both calibrated backend noise and finite-shot estimation.

\section{Discussion}\label{sec:discussion}

The results show that the compact PQC represents feasible correlated strategy distributions and achieves higher welfare than the product-strategy baselines at all evaluated PQC player counts on the Bayesian congestion game considered. The main limitation is the widening gap to the exact LP optimum, from $0.113$ at $n=4$ to $0.568$ at $n=8$, with the latter computed over the four feasible seeds. The observed gap may arise from limitations of the fixed $L=2$ ansatz, nonconvex optimization, or both; the experiments do not isolate these effects. Increasing the depth to $L=3$ at $n=4$ raises the parameter count by $50\%$ without improving mean welfare, although the variability across seeds is reduced. Thus, the gap to the LP optimum is not closed by this increase in circuit depth at the tested problem size.

Feasibility also becomes less consistent at the largest trained player count. All five seeds satisfy $\epsilon_{\max}\leq10^{-3}$ at $n=2$, $4$, and $6$, whereas four of five satisfy the tolerance at $n=8$. Because training penalizes the sum of positive aggregated obedience violations while feasibility is determined by their maximum, a reduction in the aggregate penalty does not imply that every individual obedience constraint satisfies the prescribed tolerance.

The noisy-backend simulations further highlight the distinction between welfare and feasibility. Welfare remains positive under noisy finite-shot evaluation, but no tested seed satisfies the $\epsilon_{\max}\leq10^{-3}$ tolerance, with $\epsilon_{\max}$ increasing by approximately $3.6$--$4.9$ orders of magnitude relative to the noiseless evaluations. Because these experiments combine calibrated device noise with finite-shot estimation, their individual contributions are not isolated. Error mitigation, noise-aware training, increased shot budgets, and direct hardware evaluation therefore remain important directions for further investigation.

Classical simulation also remains a scaling bottleneck. State-vector simulation with automatic-differentiation overhead produces a peak memory footprint of approximately $8.8$ GB for the single-step $n=10$ resource probe, while one gradient step requires approximately $1.5$ hours. These costs prevented PQC training at $n=10$ within the available computational resources. On quantum hardware, explicit classical state-vector storage would not be required, although execution cost would instead depend on circuit depth, connectivity, measurement requirements, device noise, and the number of circuit evaluations required by the optimization procedure.

The results are obtained under a static interaction structure, whereas strategic behavior can change substantially when interactions are network-dependent or evolve over time. Recent studies have shown that self-interaction learning, heterogeneous activation patterns, and bursty link dynamics can alter cooperation and evolutionary outcomes in networked games ~\cite{zeng2025evolutionary,zeng2025complex,zeng2025bursty}. Extending the BCE framework to these settings would require the recommendation rule and payoff model to incorporate the interaction network and its temporal evolution.

Future work should additionally examine more structured ansatzes, larger and non-binary games, and theoretical guarantees concerning expressivity, feasibility, and convergence of the penalized optimization procedure.

\section{Conclusion}\label{conclusion}

We introduced a hybrid quantum-classical framework for approximating Bayes correlated equilibrium in Bayesian games. The method uses a PQC to represent the conditional distribution over joint action profiles and optimizes expected social welfare under a penalty on positive aggregated BCE obedience violations. For an $n$-player binary-type, binary-action game, the circuit uses $2n$ qubits and $3nL$ trainable parameters, corresponding to $48$ parameters for the largest trained setting, $n=8$ with $L=2$.

On the Bayesian congestion benchmark, the PQC achieves higher welfare than the MCCFR and DCFR product-strategy baselines at every evaluated PQC player count. It satisfies the BCE feasibility criterion $\epsilon_{\max}\leq10^{-3}$ in all five runs at $n=2$, $4$, and $6$, and in four of five runs at $n=8$. The resulting welfare remains below the exact LP optimum, showing that the compact variational representation does not recover the welfare-optimal BCE over the tested range.

The computational results also distinguish representational compactness from execution cost. Although the PQC uses substantially fewer trainable parameters than an explicit type-action representation, classical state-vector simulation and gradient evaluation become increasingly expensive with larger player counts, preventing PQC training at $n=10$ within the available computational resources. The noisy-backend experiments further show that BCE feasibility is sensitive to hardware-calibrated noise and finite-shot evaluation.

These results highlight the potential of compact PQC parameterizations for approximate equilibrium computation in structured Bayesian games, while also revealing limitations in welfare optimality, feasibility at larger player counts, and classical simulation cost.

\section*{Acknowledgments}
This work was supported in part by the NYUAD Center for
Quantum and Topological Systems (CQTS), funded by Tamkeen under the NYUAD Research Institute grant CG008.

\bibliography{sn-bibliography}

\begin{appendices}

\section{Bayesian game formulation and benchmark utility}
\label{app:game_model}

This appendix provides the specification of the Bayesian game and benchmark utility used in the experiments. It distinguishes product behavioral strategies from Bayes correlated recommendation rules, defines the obedience quantities used for training and feasibility assessment, introduces the mean clipped deviation metric, and specializes the formulation to the binary-type, binary-action congestion game considered in this work.

\subsection{Bayesian game model}

A finite Bayesian game of incomplete information is defined by
\begin{equation}
\mathcal{G}
=
\left(
N,
\{\Theta_i\}_{i\in N},
\{A_i\}_{i\in N},
\pi,
\{u_i\}_{i\in N}
\right),
\end{equation}
where $N=\{1,\ldots,n\}$ is the set of players, $\Theta_i$ is the finite private type space of player $i$, and $A_i$ is the finite action space of player $i$. The joint type and action spaces are
\begin{equation}
\Theta=\prod_{i=1}^{n}\Theta_i,
\qquad
A=\prod_{i=1}^{n}A_i .
\end{equation}

The common prior over type profiles is denoted by $\pi:\Theta\rightarrow[0,1]$ and satisfies
\begin{equation}
\sum_{\theta\in\Theta}\pi(\theta)=1.
\end{equation}

Each player has a utility function
\begin{equation}
u_i:\Theta\times A\rightarrow\mathbb{R},
\end{equation}
which may depend on the type profile $\theta=(\theta_1,\ldots,\theta_n)$ and action profile $a=(a_1,\ldots,a_n)$.

Each player observes its own type before acting. A behavioral strategy for player $i$ is a mapping
\begin{equation}
\sigma_i:\Theta_i\rightarrow\Delta(A_i),
\end{equation}
and a product behavioral strategy profile induces
\begin{equation}
P_{\sigma}(a\mid\theta)
=
\prod_{i=1}^{n}
\sigma_i(a_i\mid\theta_i).
\end{equation}

Under this representation, players' actions are conditionally independent given the type profile.
Bayes correlated equilibrium allows a more general conditional recommendation rule. A mediator observes the realized type profile $\theta$ and selects a joint action recommendation according to
\begin{equation}
\sigma(a\mid\theta)\in\Delta(A),
\qquad
\sum_{a\in A}\sigma(a\mid\theta)=1.
\end{equation}

The recommended actions may therefore be correlated across players. The PQC parameterizes this conditional joint distribution as $\sigma_\phi(a\mid\theta)$.

\subsection{BCE obedience, welfare, and evaluation quantities}

For player $i$, private type $\theta_i$, recommended action $a_i$, and alternative action $a'_i$, we define the aggregated BCE obedience quantity
\begin{equation}
\begin{aligned}
V_{i,\theta_i,a_i,a'_i}(\sigma)
&=
\sum_{\theta_{-i},a_{-i}}
\pi(\theta_i,\theta_{-i})
\sigma(a_i,a_{-i}\mid\theta_i,\theta_{-i}) \\
&\quad\times
\left[
u_i\bigl((\theta_i,\theta_{-i}),(a'_i,a_{-i})\bigr)
-
u_i\bigl((\theta_i,\theta_{-i}),(a_i,a_{-i})\bigr)
\right].
\end{aligned}
\label{eq:app_bce_violation}
\end{equation}

A Bayes correlated recommendation rule satisfies the obedience conditions when
\begin{equation}
V_{i,\theta_i,a_i,a'_i}(\sigma)\leq0
\end{equation}
for every player, private type, recommended action, and unilateral alternative.

Equation~\eqref{eq:app_bce_violation} uses the joint prior directly and is not normalized by the probability of observing $(\theta_i,a_i)$. When this event has positive probability, normalization multiplies the expression by a positive constant and therefore does not change the sign of the obedience condition. When the event has zero probability, the corresponding unnormalized condition is zero and hence vacuous. This unnormalized form is used consistently in the PQC objective, the evaluation procedure, and the exact LP reference.

In the binary-action setting considered here, each recommended action has a unique unilateral alternative, $a'_i=1-a_i$. The positive obedience penalty used during PQC training is
\begin{equation}
P(\phi)
=
\sum_{i=1}^{n}
\sum_{\tau\in\{0,1\}}
\sum_{a_i\in\{0,1\}}
\left[
V_{i,\tau,a_i,1-a_i}(\sigma_\phi)
\right]_+,
\label{eq:app_penalty}
\end{equation}
where
\begin{equation}
[x]_+=\max(0,x).
\end{equation}

BCE feasibility is assessed using the maximum positive aggregated obedience violation,
\begin{equation}
\epsilon_{\max}(\phi)
=
\max_{i,\tau,a_i}
\left[
V_{i,\tau,a_i,1-a_i}(\sigma_\phi)
\right]_+.
\label{eq:app_epsilon}
\end{equation}

A learned strategy is classified as feasible when
\begin{equation}
\epsilon_{\max}(\phi)\leq10^{-3}.
\end{equation}

The social welfare of a type-action profile is
\begin{equation}
W(\theta,a)
=
\sum_{i=1}^{n}u_i(\theta,a),
\end{equation}

and its expectation under the PQC-induced recommendation rule is
\begin{equation}
\overline{W}(\phi)
=
\sum_{\theta\in\Theta}
\pi(\theta)
\sum_{a\in A}
\sigma_\phi(a\mid\theta)
W(\theta,a).
\label{eq:app_welfare}
\end{equation}

The training objective is
\begin{equation}
\mathcal{L}_t(\phi)
=
-\overline{W}(\phi)
+
\lambda_t P(\phi),
\label{eq:app_loss}
\end{equation}
with penalty weight
\begin{equation}
\lambda_t
=
\min\left(50,1.03^t\right).
\end{equation}

The circuit is therefore optimized to increase expected welfare while progressively penalizing positive aggregated BCE obedience violations.
For evaluation, we define the gain from flipping the current binary action as
\begin{equation}
g_i(\theta,a)
=
u_i\bigl(\theta,1-a_i,a_{-i}\bigr)
-
u_i\bigl(\theta,a_i,a_{-i}\bigr),
\label{eq:app_deviation_gain}
\end{equation}

and the mean clipped deviation metric
\begin{equation}
R(\phi)
=
\frac{1}{n}
\sum_{i=1}^{n}
\sum_{\theta\in\Theta}
\pi(\theta)
\sum_{a\in A}
\sigma_\phi(a\mid\theta)
\left[g_i(\theta,a)\right]_+.
\label{eq:app_R}
\end{equation}

Unlike $P(\phi)$ and $\epsilon_{\max}(\phi)$, $R(\phi)$ clips each realized deviation gain before aggregation over type and action profiles. It is used only as an evaluation metric and is neither the PQC training objective nor the BCE feasibility criterion.

The two quantities are related but are not equivalent. Using
\begin{equation}
\left[\sum_j x_j\right]_+
\leq
\sum_j[x_j]_+,
\end{equation}
and the nonnegativity of $\pi(\theta)$ and $\sigma_\phi(a\mid\theta)$, any obedience constraint satisfies
\begin{equation}
\begin{aligned}
\left[
V_{i,\tau,a_i,1-a_i}(\sigma_\phi)
\right]_+
&\leq
\sum_{\substack{\theta\in\Theta:\theta_i=\tau}}
\pi(\theta)
\sum_{a_{-i}}
\sigma_\phi(a_i,a_{-i}\mid\theta)
\left[
g_i\bigl(\theta,(a_i,a_{-i})\bigr)
\right]_+ \\
&\leq
\sum_{\theta\in\Theta}
\pi(\theta)
\sum_{a\in A}
\sigma_\phi(a\mid\theta)
\left[g_i(\theta,a)\right]_+ .
\end{aligned}
\end{equation}

Consequently,
\begin{equation}
\epsilon_{\max}(\phi)
\leq
nR(\phi).
\label{eq:app_eps_R_relation}
\end{equation}

Thus, $R(\phi)=0$ implies $\epsilon_{\max}(\phi)=0$ and certifies exact BCE obedience. The converse does not necessarily hold: positive pointwise deviation gains can be offset by negative gains within an aggregated obedience constraint, allowing $\epsilon_{\max}(\phi)=0$ while $R(\phi)>0$.

\subsection{Binary-type and binary-action specialization}

The experiments use
\begin{equation}
\Theta_i=\{0,1\},
\qquad
A_i=\{0,1\},
\qquad
i=1,\ldots,n,
\end{equation}
with independent and uniformly distributed types. Hence,
\begin{equation}
\pi(\theta)=2^{-n},
\qquad
\theta\in\{0,1\}^{n},
\end{equation}
and
\begin{equation}
|\Theta|=2^n,
\qquad
|A|=2^n.
\end{equation}

An explicit conditional recommendation table therefore contains
\begin{equation}
|\Theta||A|
=
2^n2^n
=
4^n
\end{equation}
probability variables.

Because each player has two possible types, two possible recommended actions, and exactly one unilateral alternative for each recommended action, there are
\begin{equation}
n\times2\times2
=
4n
\end{equation}
nontrivial BCE obedience inequalities. The exact LP reference solves
\begin{equation}
\max_{\sigma}\;
\sum_{\theta\in\Theta}
\pi(\theta)
\sum_{a\in A}
\sigma(a\mid\theta)W(\theta,a)
\end{equation}
subject to
\begin{equation}
V_{i,\tau,a_i,1-a_i}(\sigma)\leq0
\qquad
\forall i,\tau,a_i,
\end{equation}
with the additional constraints
\begin{equation}
\sum_{a\in A}\sigma(a\mid\theta)=1
\qquad
\forall\theta\in\Theta,
\end{equation}
and
\begin{equation}
\sigma(a\mid\theta)\geq0
\qquad
\forall\theta,a.
\end{equation}

At $n=10$, the joint type space and joint action space each contain
\begin{equation}
2^{10}=1024
\end{equation}
profiles. The explicit LP representation therefore contains
\begin{equation}
4^{10}
=
2^{20}
=
1{,}048{,}576
\end{equation}
type-action probability variables, along with $40$ nontrivial BCE obedience inequalities and $1024$ conditional normalization constraints.

\subsection{Bayesian congestion benchmark utility}

The experiments use a Bayesian congestion game with two available resources. For an action profile $a$, we define
\begin{equation}
m_x(a)
=
\sum_{j=1}^{n}\mathbf{1}\{a_j=x\},
\qquad
x\in\{0,1\},
\end{equation}

where $m_x(a)$ denotes the number of players selecting resource $x$. The utility of player $i$ is
\begin{equation}
u_i(\theta,a)
=
1
+
\beta\,\theta_i a_i
-
\kappa\,\frac{m_{a_i}(a)}{n},
\label{eq:app_congestion_payoff}
\end{equation}

with
\begin{equation}
\beta=0.5,
\qquad
\kappa=2.
\end{equation}

The term $\beta\theta_i a_i$ provides a type-dependent bonus to a type-$1$ player choosing action $1$, while the congestion term reduces utility as more players select the same resource. The payoff therefore depends directly on the joint action profile.

The corresponding social welfare is obtained by summing Eq.~\eqref{eq:app_congestion_payoff} over all players:
\begin{equation}
W(\theta,a)
=
n
+
\beta\sum_{i=1}^{n}\theta_i a_i
-
\frac{\kappa}{n}
\sum_{i=1}^{n}m_{a_i}(a).
\end{equation}

For each resource $x$, exactly $m_x(a)$ players incur the congestion term $m_x(a)$, so
\begin{equation}
\sum_{i=1}^{n}m_{a_i}(a)
=
m_0(a)^2+m_1(a)^2.
\end{equation}

Hence,
\begin{equation}
W(\theta,a)
=
n
+
\beta\sum_{i=1}^{n}\theta_i a_i
-
\frac{\kappa}{n}
\left[
m_0(a)^2+m_1(a)^2
\right].
\label{eq:app_congestion_welfare}
\end{equation}

The first term involving $\theta_i a_i$ favors assigning type-$1$ players to action $1$, whereas the quadratic congestion term favors distributing players across the two resources.

The dependence of unilateral incentives on the actions of the other players can be derived explicitly. Let
\begin{equation}
s_i
=
\sum_{j\neq i}a_j
\end{equation}

denote the number of opponents choosing action $1$. If player $i$ chooses action $0$, the number of players using resource $0$ is $n-s_i$, whereas after switching to action $1$, the number using resource $1$ is $s_i+1$. The gain from switching from $0$ to $1$ is therefore
\begin{align}
\Delta_i(\theta_i,s_i)
&=
u_i\bigl(\theta,1,a_{-i}\bigr)
-
u_i\bigl(\theta,0,a_{-i}\bigr) \\
&=
\left(
1+\beta\theta_i
-\kappa\frac{s_i+1}{n}
\right)
-
\left(
1-\kappa\frac{n-s_i}{n}
\right) \\
&=
\beta\theta_i
+
\frac{\kappa}{n}
\left(
n-1-2s_i
\right).
\label{eq:app_congestion_deviation}
\end{align}

The gain from flipping the currently selected action is then given by
\begin{equation}
g_i(\theta,a)
=
(1-2a_i)
\left[
\beta\theta_i
+
\frac{\kappa}{n}
\left(
n-1-2\sum_{j\neq i}a_j
\right)
\right].
\label{eq:app_flip_gain_closed}
\end{equation}

For $a_i=0$, Eq.~\eqref{eq:app_flip_gain_closed} gives the $0\rightarrow1$ deviation gain in Eq.~\eqref{eq:app_congestion_deviation}; for $a_i=1$, the sign reverses and gives the $1\rightarrow0$ deviation gain. Thus, unilateral incentives depend explicitly on the actions of the other players, establishing the strategic coupling in the benchmark.

\end{appendices}


\end{document}